\documentclass[twoside]{ilcws10}
\usepackage[latin1]{inputenc}
\usepackage[dvips]{graphicx,epsfig,color}
\usepackage{wrapfig,rotating}
\usepackage{amssymb,amsmath,array}

\begin{document}
\title{
Monte Carlo Studies of the CALICE AHCAL Tiles Gaps and Non-uniformities} 
\author{Angela Lucaci-Timoce$^1$ and Felix Sefkow$^2$
\vspace{.3cm}\\
1- DESY Hamburg\\
Bld. 1, Rm. O1.549\\
Notkestr. 85 - Germany
\vspace{.1cm}\\
2- DESY Hamburg \\
Bld. 1, Rm. O1.515\\
Notkestr. 85 - Germany
}

\maketitle

\begin{abstract}
The CALICE analog HCAL is a highly granular calorimeter, proposed for the International Linear Collider. It is based on scintillator tiles, read out by silicon photomultipliers (SiPMs). The effects of gaps between the calorimeter tiles, as well as the non-uniform response of the tiles, in view of the impact on the energy resolution, are studied in Monte Carlo events. It is shown that these type of effects do not have a significant influence on the measurement of hadron showers.
\end{abstract}

\section{Introduction}

\begin{wrapfigure}{r}{0.5\columnwidth}
\centerline{\includegraphics[scale=0.2]{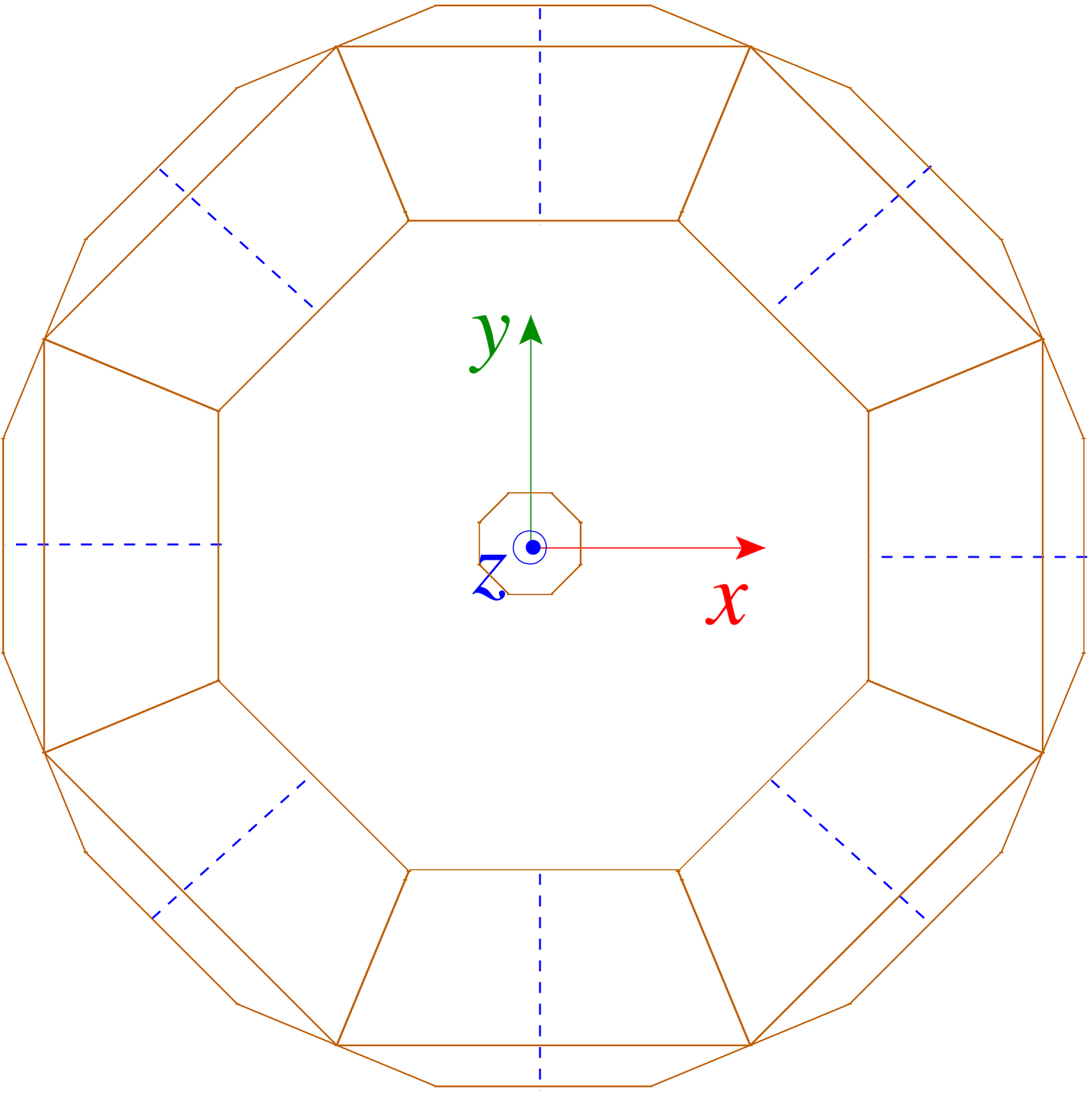}}
\caption{Octahedral shape of the AHCAL.}
\label{fig:HCAL-staves}
\end{wrapfigure}

The HCAL is designed to have an octahedral structure, consisting of eight identical staves, each divided in two halves in the $(x,\;y)$ plane, as shown in Fig.~\ref{fig:HCAL-staves}. Along the $z$-axis, the calorimeter is again divided in two modules, with a gap of 2 mm. Additionally, the layers which contain the scintillator tiles have a support structure of 5 mm. This design is currently implemented in Mokka, which is a GEANT4 application used for the ILC detector simulation. A complete description of the scintillator HCAL as implemented in Mokka can be found in \cite{Hcal-in-Mokka-ILD-note}.

The front-end electronics will be integrated into the absorber structure, whereas the electronics connections and interfaces will be placed at the 2 end-faces, for easy access for maintenance and service lines.  To keep the single electronics modules at reasonable size, the detector's electronics is divided into basic units, each with a typical size of $36\;\times\;36\;\textrm{cm}^2$, integrating 144 tiles, together with the corresponding SiPMs, front-end electronics and the light calibration system. The analogue signals of the SiPMs are read out by 4 front-end ASICs of type SPIROC, developed by LAL/Omega. For more details about the AHCAL engineering prototype, see \cite{EUDET-report}.

Although the description of the scintillator HCAL in Mokka is highly detailed, it does not contain gaps between the scintillator tiles, as it would be in a real calorimeter. Also no gaps between the electronics units are simulated, as well as no tiles non-uniformities.  All these are, in principle, expected to influence negatively the calorimeter energy resolution, as presented for example by the CALICE Si-W ECAL group in~\cite{ECAL-paper}. However, while this is true for the narrow electromagnetic showers, the situation is different for hadron showers, as it will be shown in the next sections.

\section{Simulation of the gaps between the tiles and between the electronic units}


For studying the effect of gaps, a new Mokka driver for the HCAL was developed, similar to the test beam HCAL, with 38 layers, each with 20 mm Fe absorber, but with $3\times3\;\textrm{cm}^2$ tiles. In addition, in each layer, $6\times 6$ tiles were grouped together to simulate the HCAL base electronic unit (HBU).

\begin{wraptable}{l}{0.5\columnwidth}
\centerline{\begin{tabular}{|l|l|l|}
\hline
Gaps  & between tiles & between HBUs \\\hline
\textbf{unrealistic}  & 1.5 mm & 5 mm  \\
\textbf{realistic} & 0.15 mm & 0.5 mm\\\hline
\end{tabular}}
\caption{Gaps implemented between the scintillator tiles and the HCAL base electronic units (HBUs).}
\label{tab:gaps}
\end{wraptable}

 Two types of gaps were implemented, as presented in Table~\ref{tab:gaps}. The realistic gaps are corresponding to the ones in the actually built prototype model, whereas the unrealistic values are introduced in order to verify that the gap effects are in principle properly modeled by the simulation.

Note that the studies were done with simulated hits only, without any digitisation. Two types of particles were used: $e^-$ and $\pi^-$, each with 5 and 50 GeV beam energies, respectively, and without a beam spread. A beam position scan of 30 points in $x$ and in $y$ was performed, and for each possible combination 10000 events were generated.

\subsection{Tiles with wavelength shifting fiber}
\begin{wrapfigure}{r}{0.5\columnwidth}
\centerline{\includegraphics[scale=0.065]{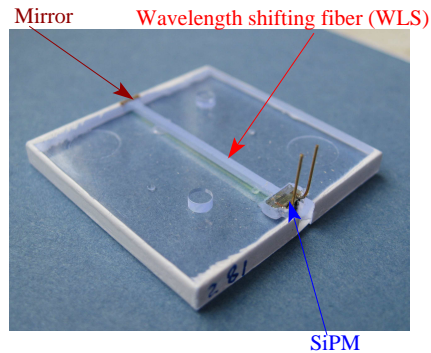}}
\caption{Example of a scintillator tile with a wavelength shifting fiber in the center, coupled to a SiPM}
\label{fig:scintillatorTile}
\end{wrapfigure}

A possible realisation of a scintillator tile with a wavelength shifting fiber is presented in Fig.~\ref{fig:scintillatorTile}. Measurements with such tiles were performed at the ITEP institute (Russia) in a hadron test beam, using a wire chamber tracking system with a resolution of approximately 1 mm. It was shown that the left and right edges of the tiles have around 10\% less response than the rest of the tile due to a 100 $\mu$m wide zone with zero response, whereas in the fiber region 20\% less response was registered. The lowest response was measured in the region corresponding to the mirror and to the SiPM cut-outs. 

Therefore,  these \textbf{dead regions} (no signal in the mirror and the SiPM cut-outs, and 20\% less response in the fiber region) were also considered in the simulation.

\begin{figure}[thb]
\begin{center}
\includegraphics[scale=0.25]{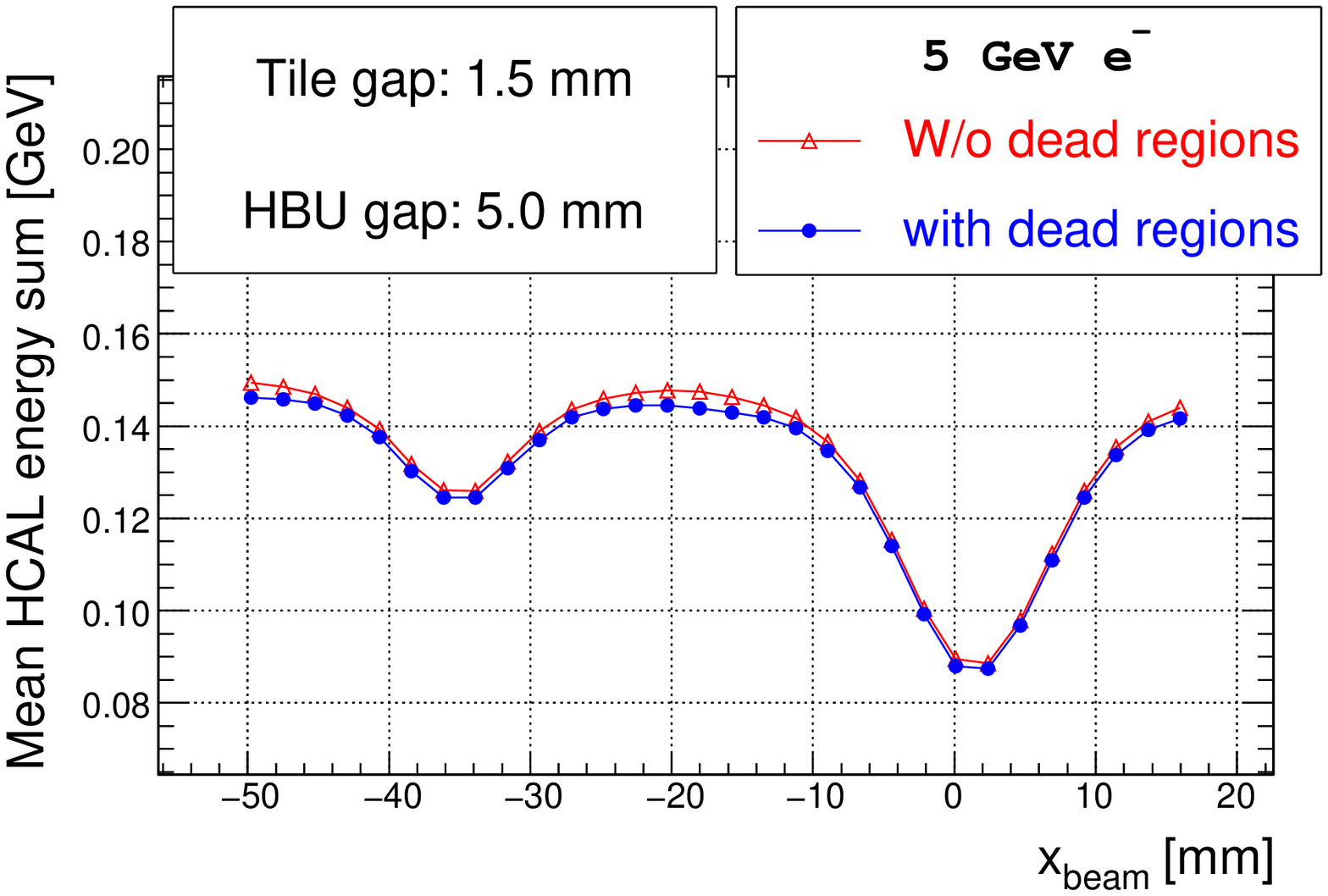}
\includegraphics[scale=0.25]{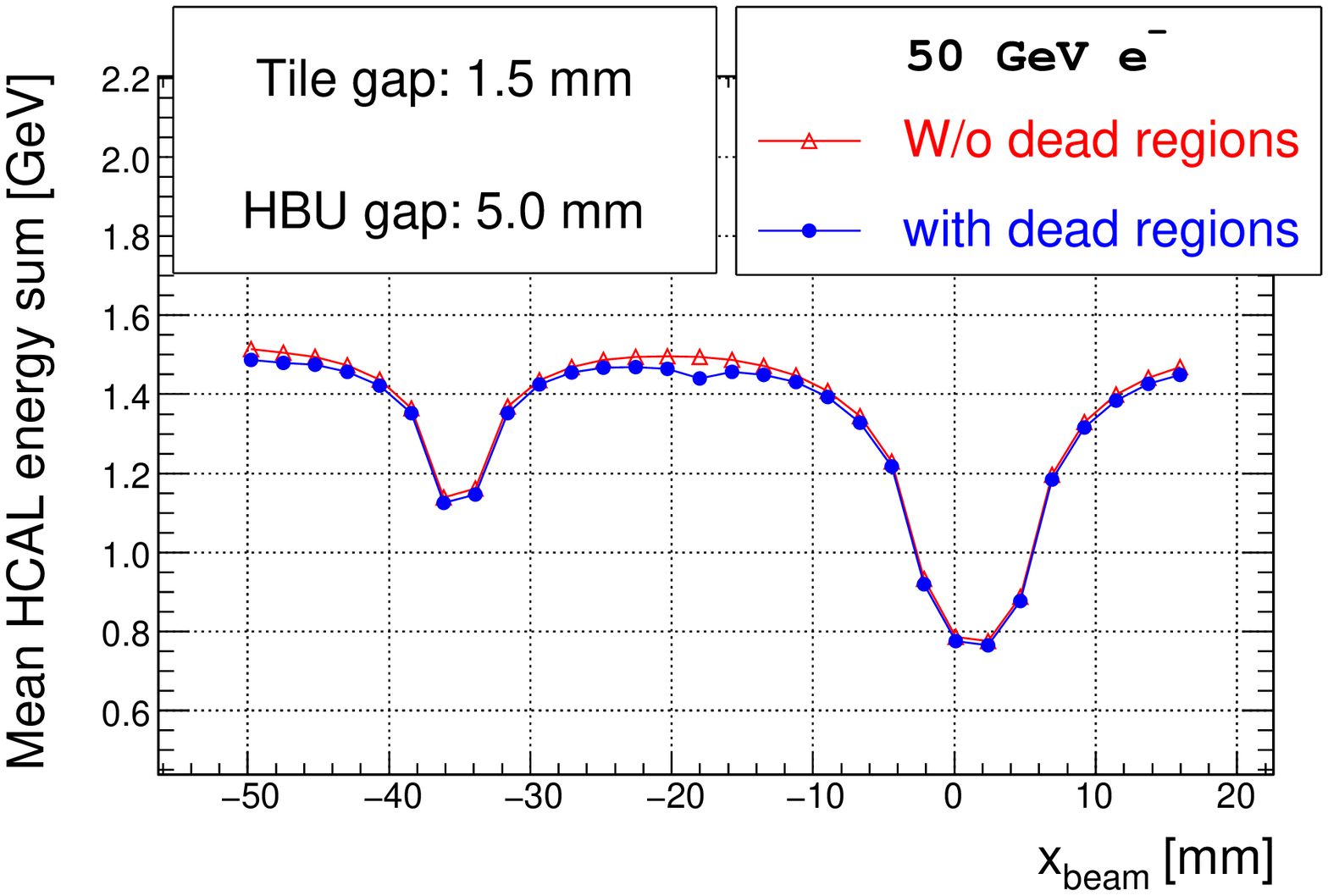}\\
\includegraphics[scale=0.25]{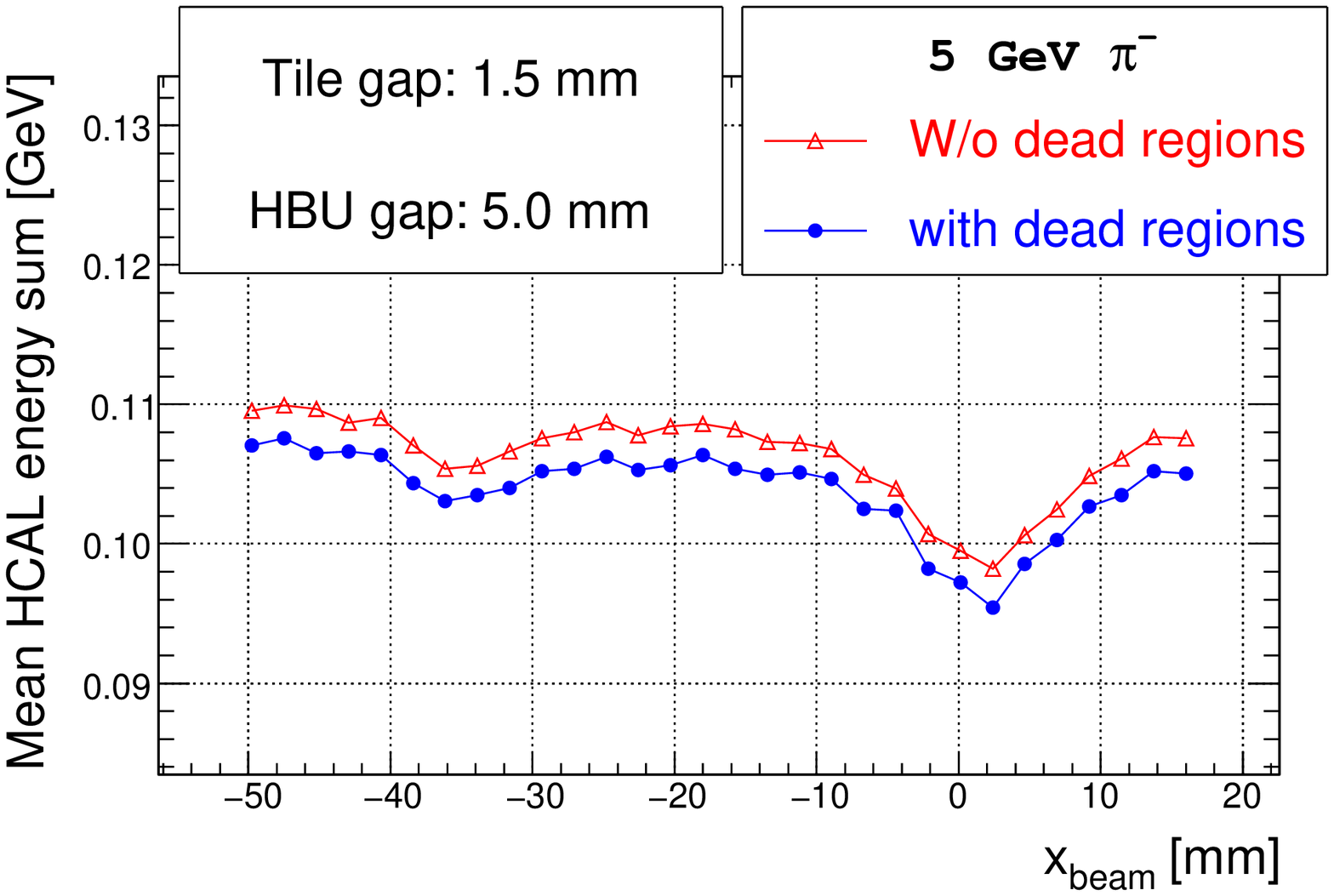}
\includegraphics[scale=0.25]{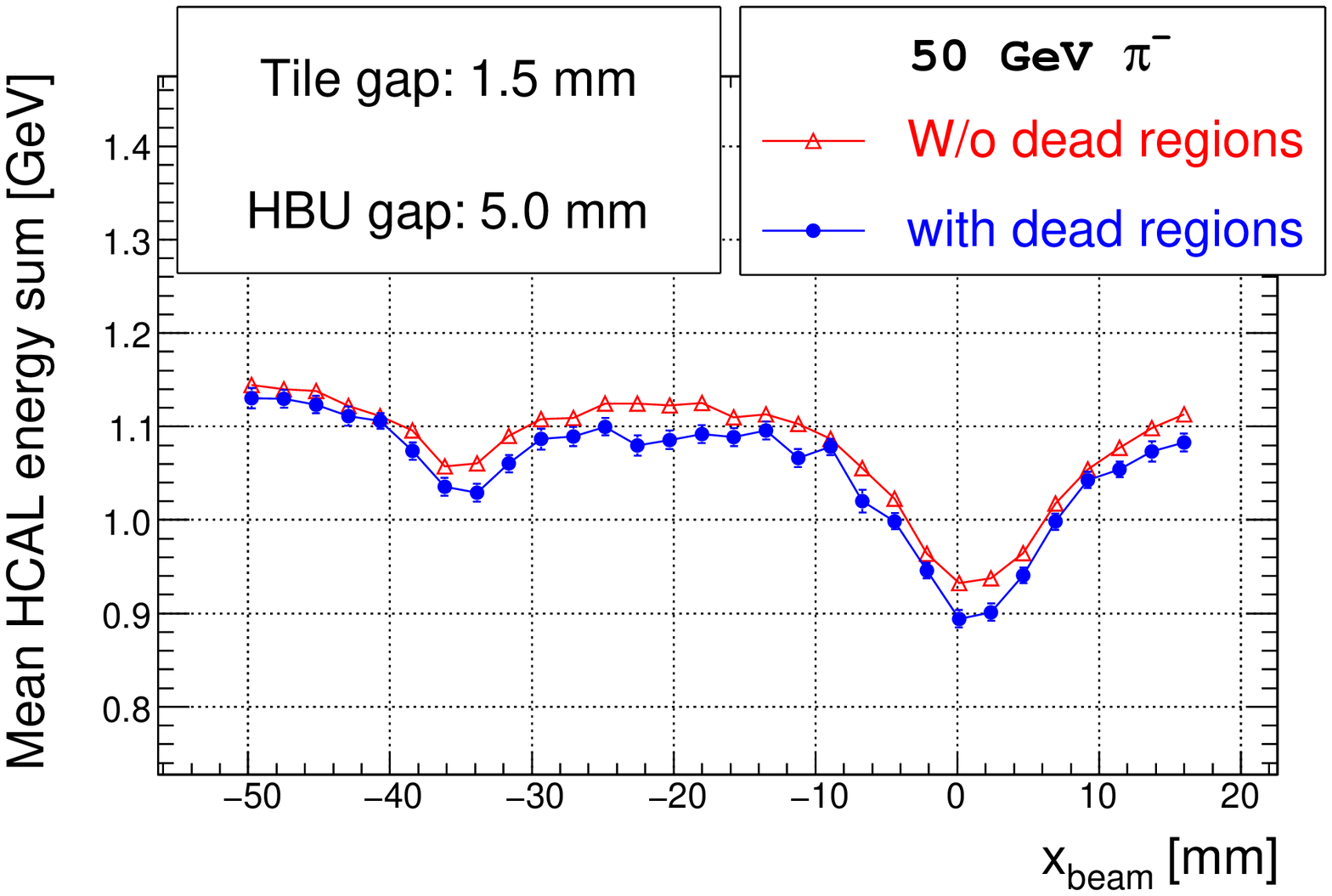}\\
\caption{Mean HCAL energy sum as a function of the beam position along the $x$-axis, in case of \textbf{unrealistic} gaps between tiles and between HBUs for electrons (top) and pions (bottom).}
\end{center}
\label{fig:gaps-unrealistic}
\end{figure}

\begin{figure}[h!]
\begin{center}
\includegraphics[scale=0.25]{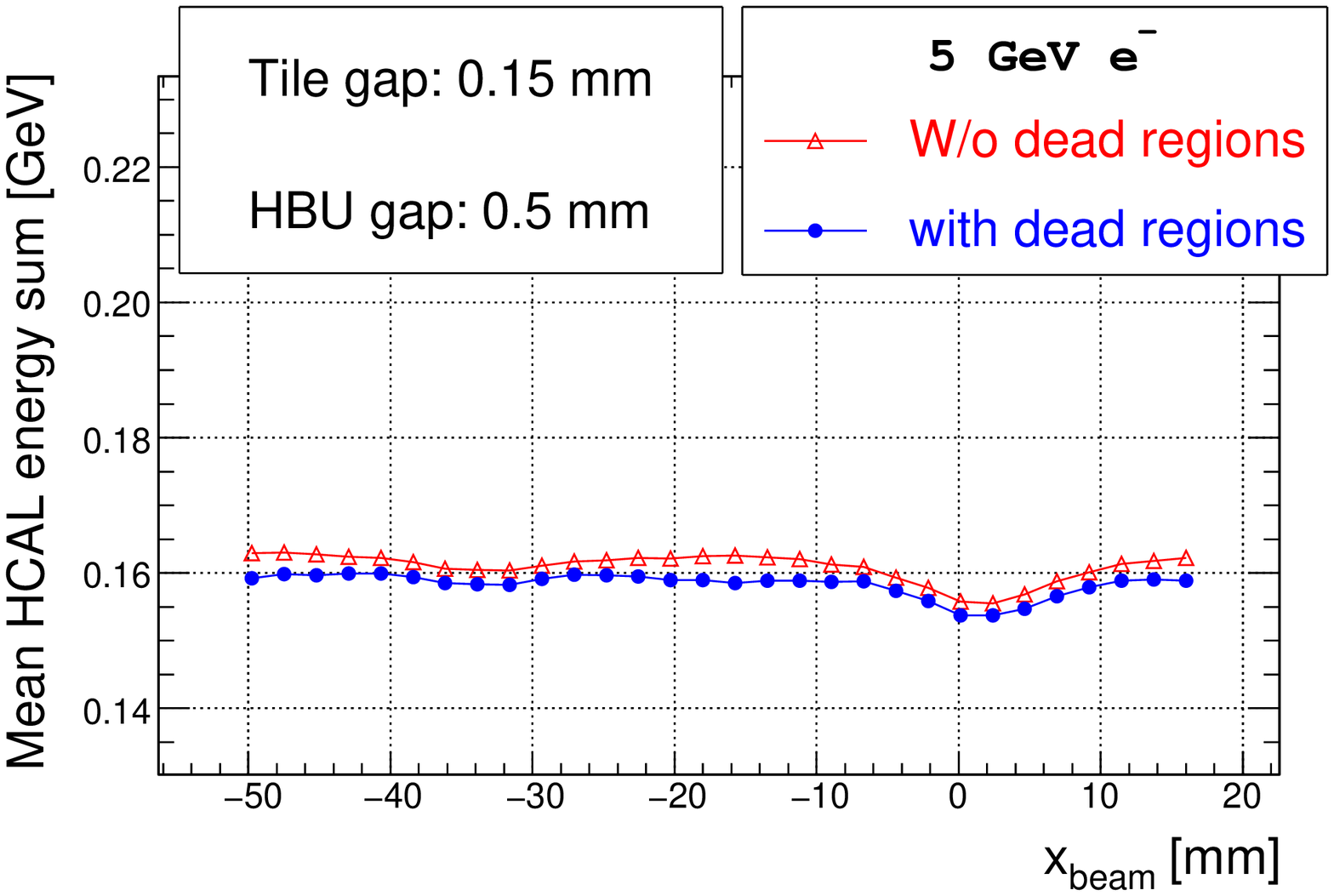}
\includegraphics[scale=0.25]{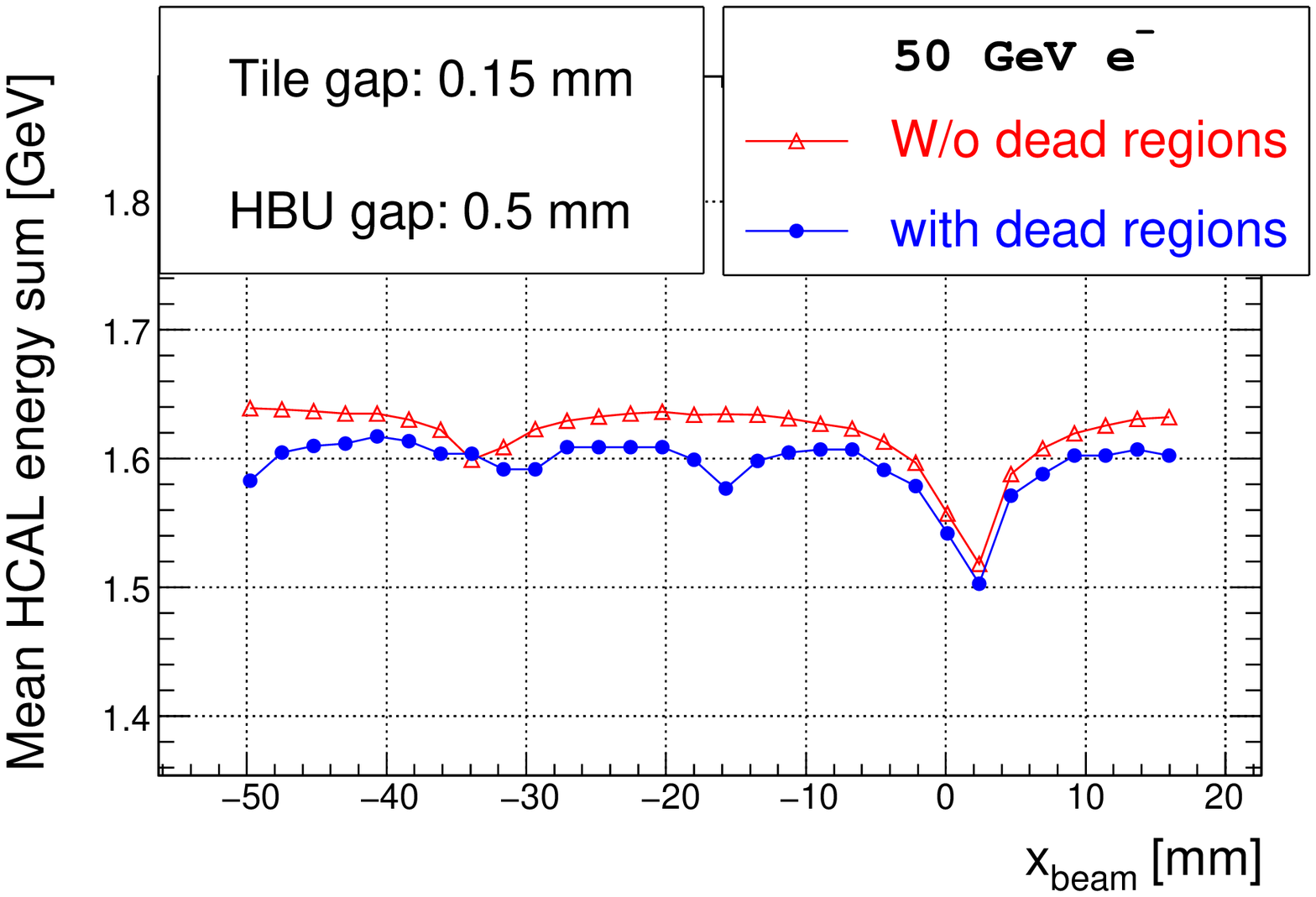}\\
\includegraphics[scale=0.25]{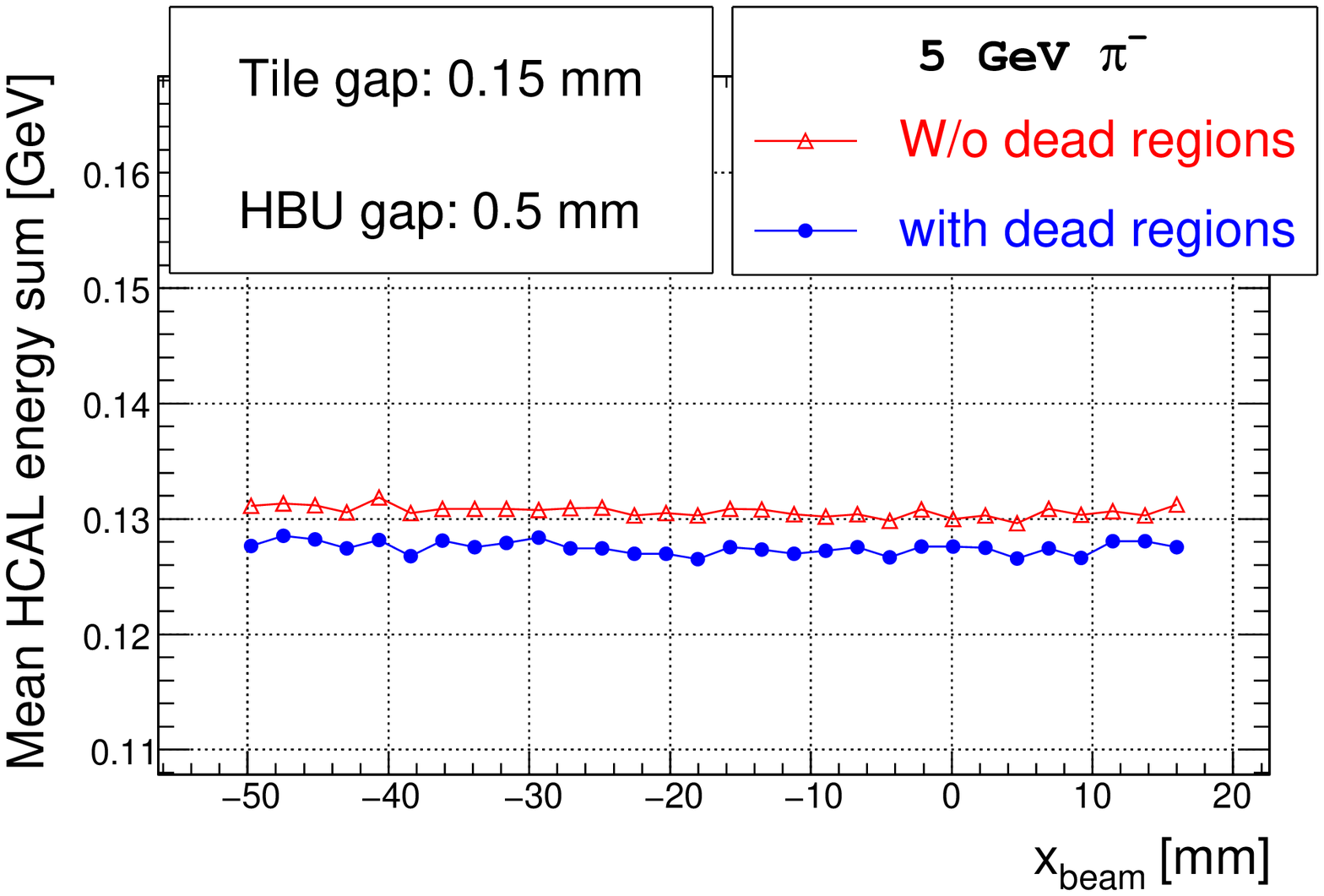}
\includegraphics[scale=0.25]{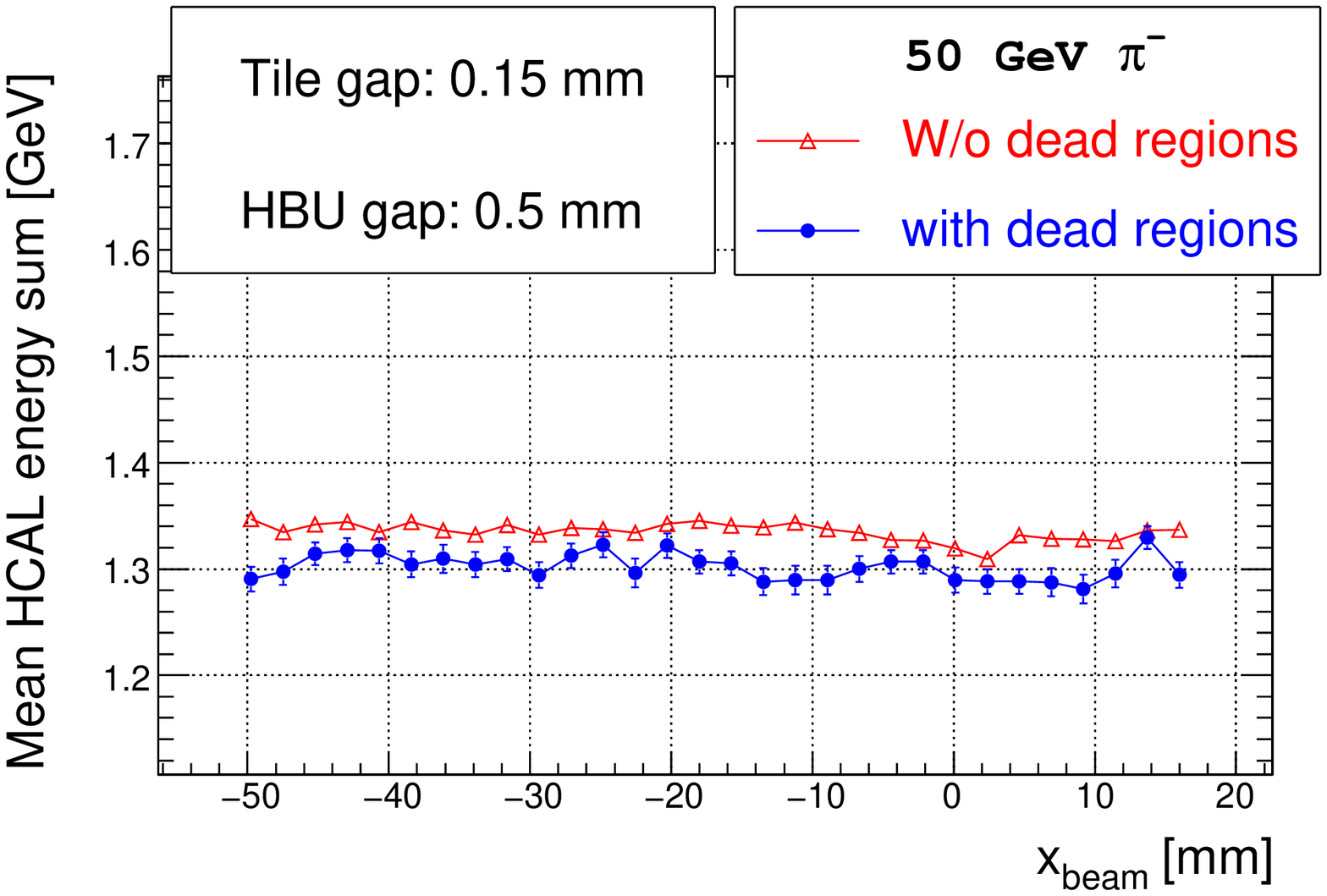}\\
\caption{Mean HCAL energy sum as a function of the beam position along the $x$-axis, in case of \textbf{realistic} gaps between tiles and between HBUs for electrons (top) and pions (bottom).}
\end{center}
\label{fig:gaps-realistic}
\end{figure}
In an initial study, the $y$-position of the beam was placed in the center of the tile, and a scan was done along the $x$-axis. The total energy deposited in the HCAL was summed up, and the arithmetic mean of the total energy sum is presented as a function of the beam position along the $x$-axis in Fig.~3 for the case with very large gaps. The first dip observed in the electron case is due to the gap between the tiles, and the second due to the gap between the electronic units. Since the hadron showers are broader than the electromagnetic ones, the dips due to gaps are less pronounced in the pion case. In Fig.~4, the same distributions are presented, but this time for realistic values.

In a second series of studies, the $y$-position of the beam was placed at the lower margin of the tile, and a scan was performed along the $x$-axis, in order to see the effect of the SiPM cut-out (not shown). As expected, the SiPM cut-out was visible in case of the narrow electromagnetic showers, but less significant in case of pion induced showers.

\subsection{Tiles with direct coupling of SiPM}
New developments of the SiPMs lead to building of devices which are blue sensitive, and thus well matched to the emission spectrum of the scintillator. As a result, the wavelength shifting fiber is not necessary anymore, which lead to a simplification of the production of the scintillator tiles.  The CALICE group at MPI  (Munich) performed extensive tests of this new type of scintillator tiles (for more details, see~\cite{Soldner}). First, they considered the simple case of the SiPM directly coupled to the tile. However, this resulted in  significant non-uniformity of the tile's response, as shown in Fig.~5, left. To improve the the uniformity, they integrated the SiPM into a deep slit into a tile (see Fig.~5, right).

\begin{figure}[h!]
\begin{center}
\includegraphics[scale=0.3]{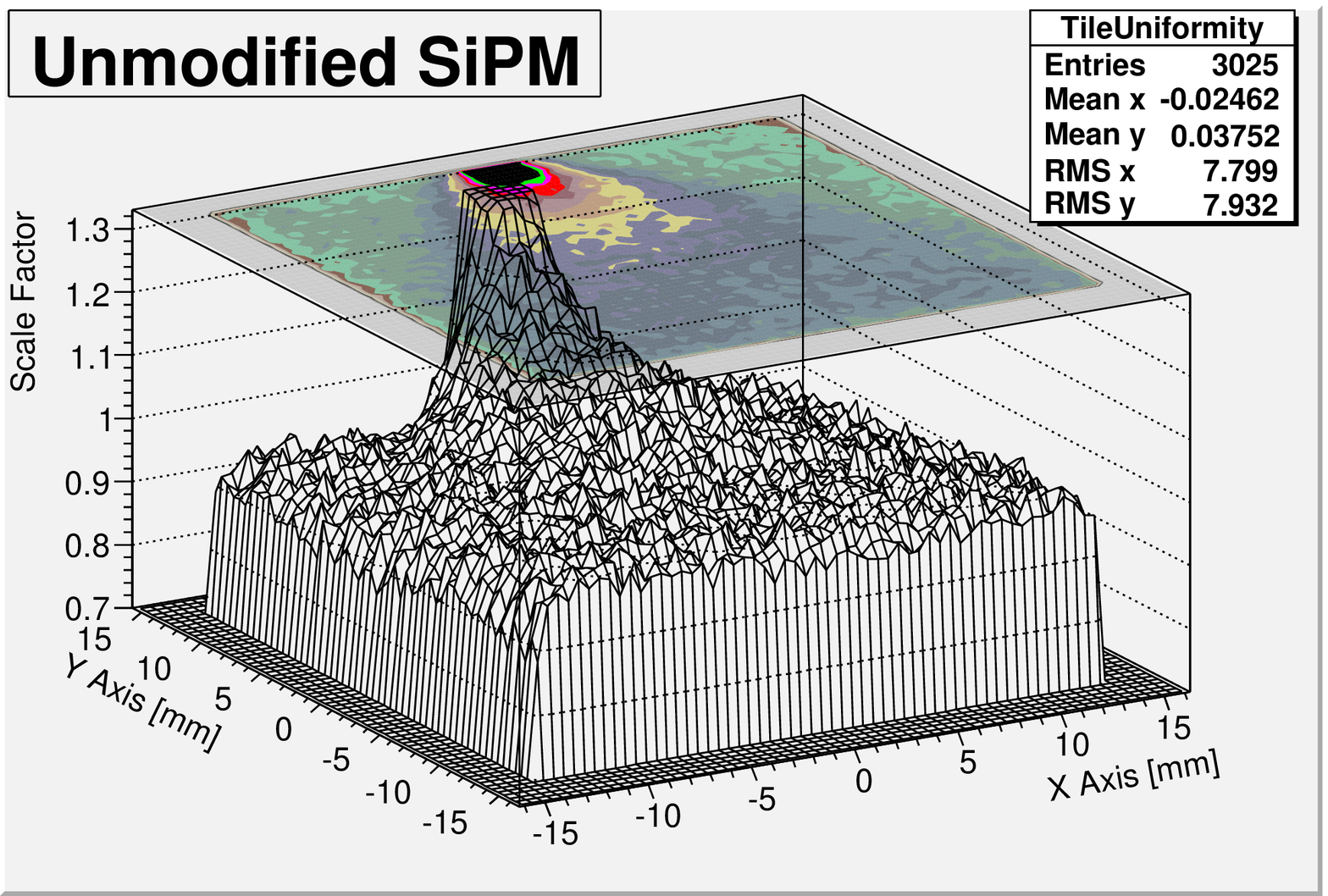}
\hspace{0.5cm}
\includegraphics[scale=0.3]{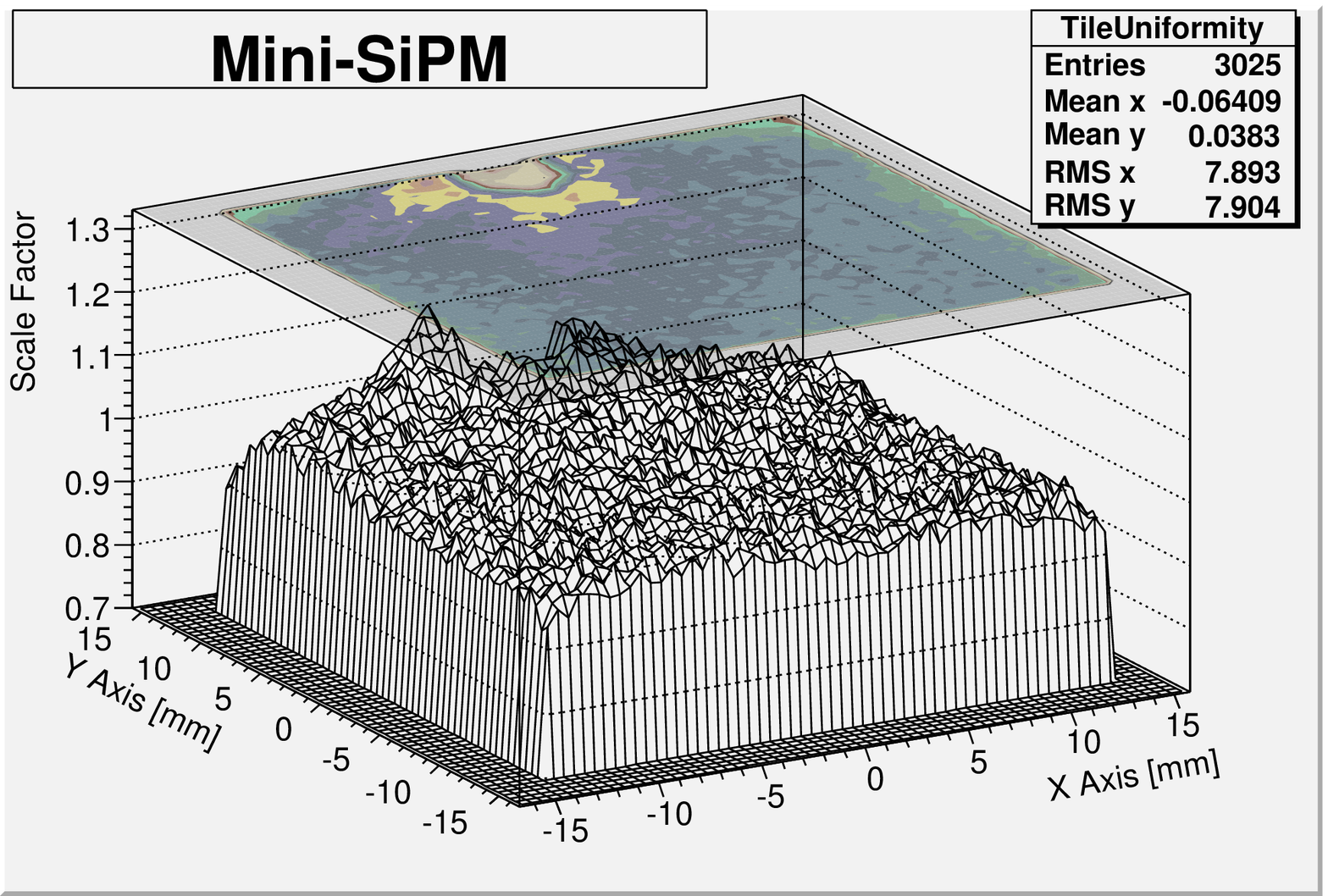}
\caption{The measured scale factor of the tile's response as a function of the $x$ and $y$ position in the tile for the case of a tile with the SiPM directly coupled to the tile (left) and with the SiPM integrated into the tile, into a deep slit (right).}
\end{center}
\label{fig:directCoupling}
\end{figure}

The measured scaling factors of the tile's response were applied to the simulated HCAL hits, and the same studies as for the tile with a wavelength shifting fiber were redone. The results are presented in Fig~6. It is clear that the non-uniformity has a visible impact in electron-induced showers, whereas for the pion case the effects are much reduced. When scanning along the $y$-axis, the peak due to the high increase of the tile's response in the SiPM region is clearly visible. Therefore, the solution to be preferred would be the tile with an integrated SiPM, which offers the best non-uniformity.

\begin{figure}[h!]
\begin{center}
\includegraphics[scale=0.3]{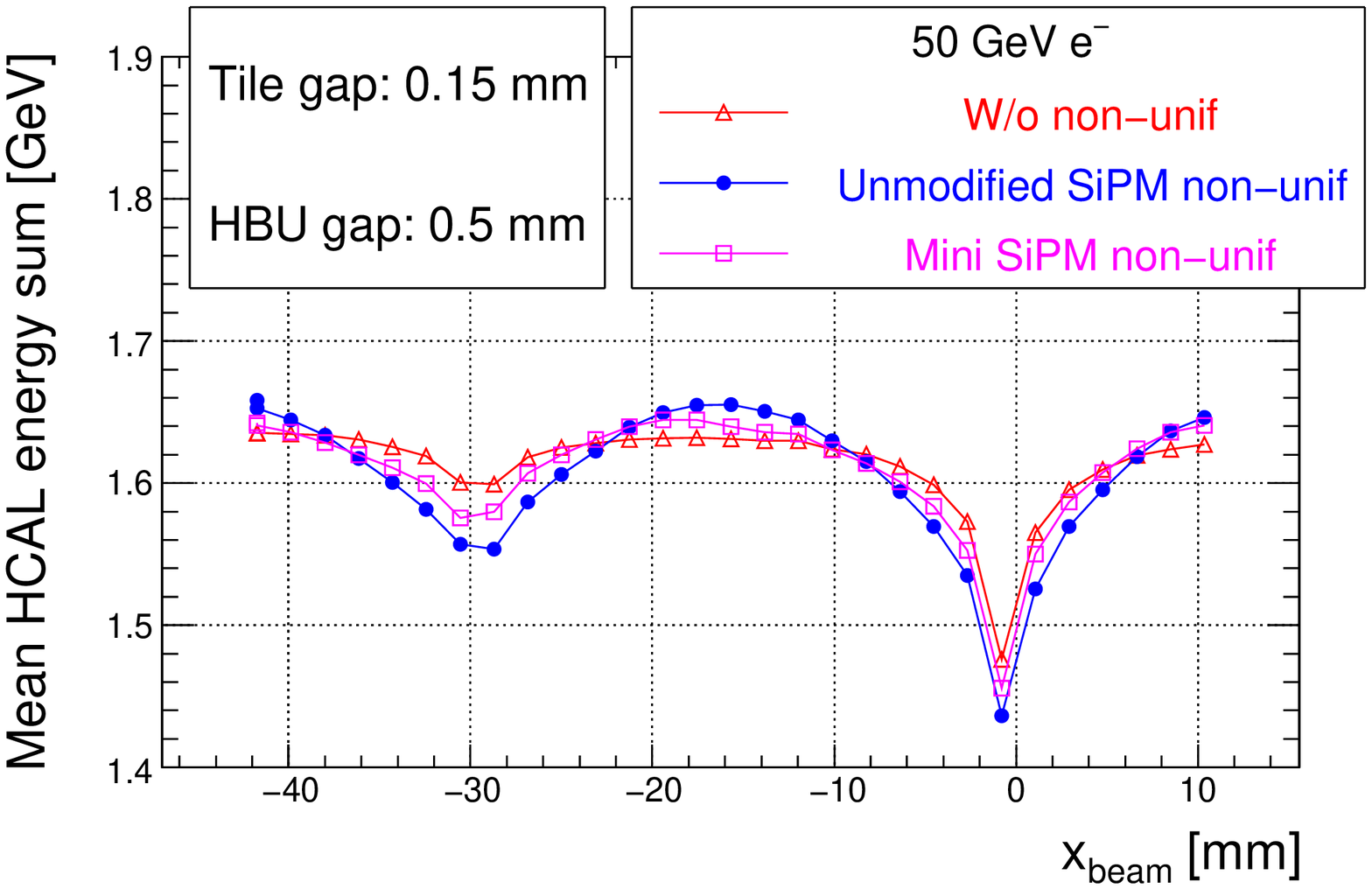}
\includegraphics[scale=0.3]{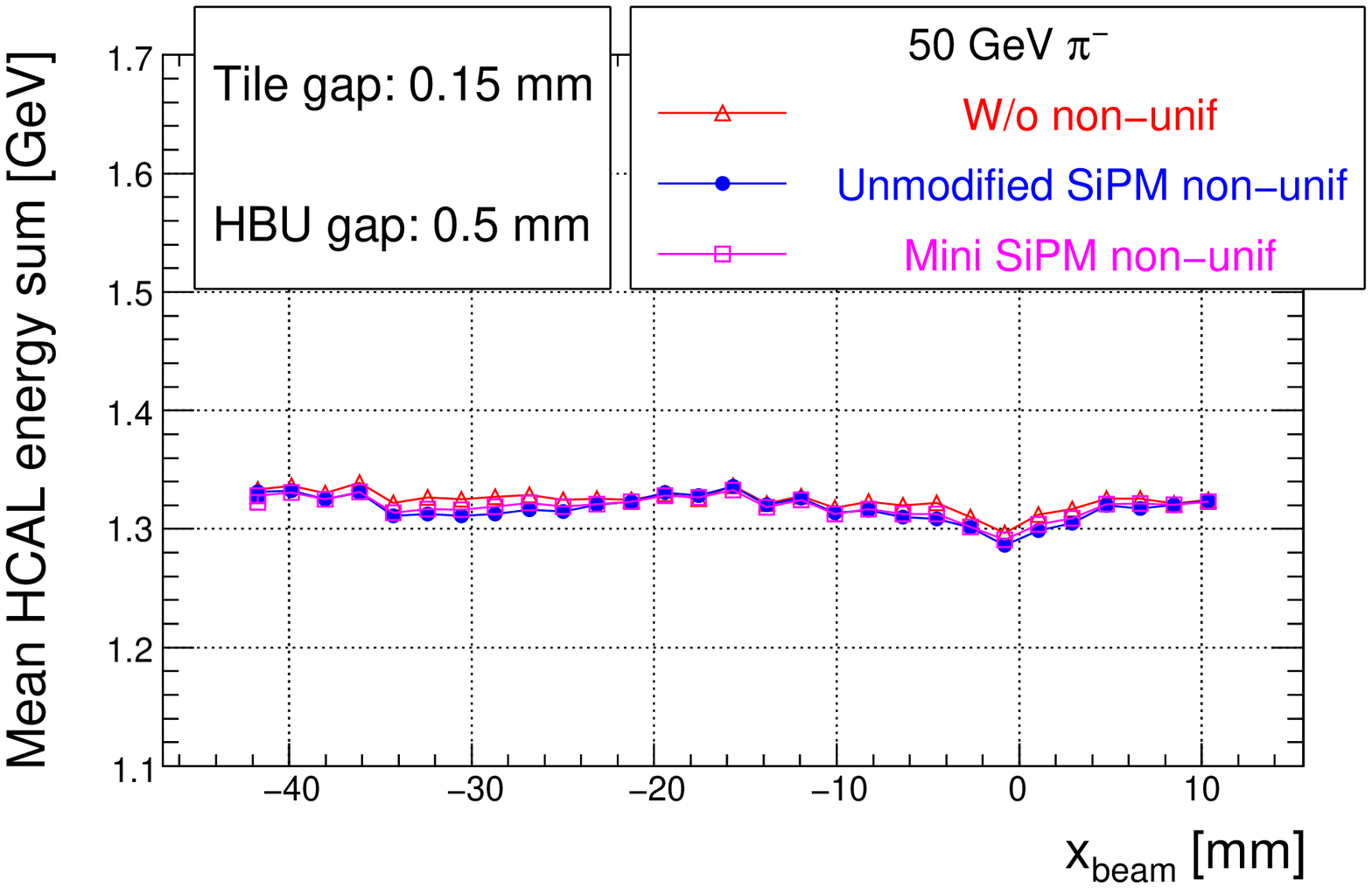}\\
\includegraphics[scale=0.3]{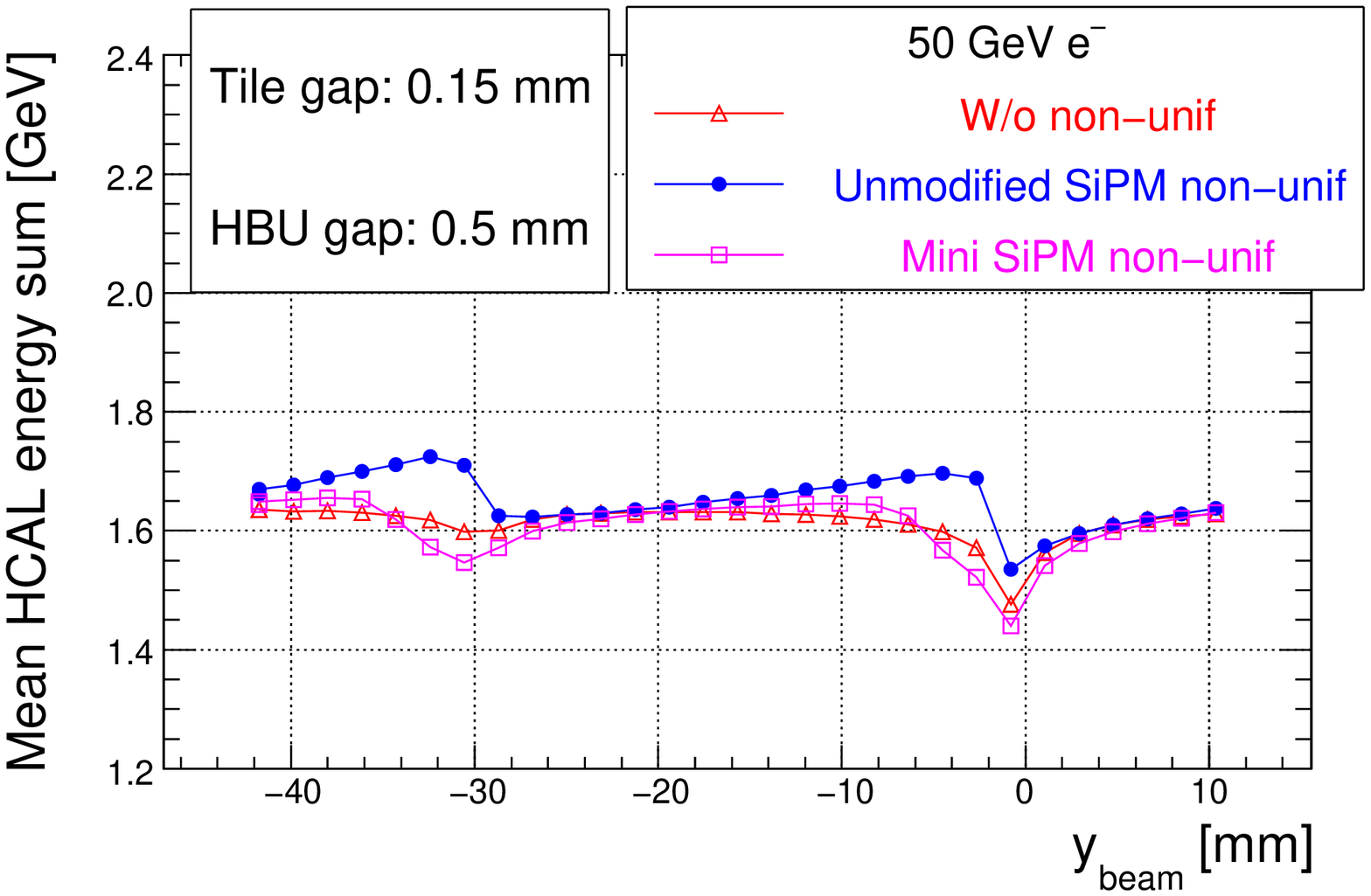}
\includegraphics[scale=0.3]{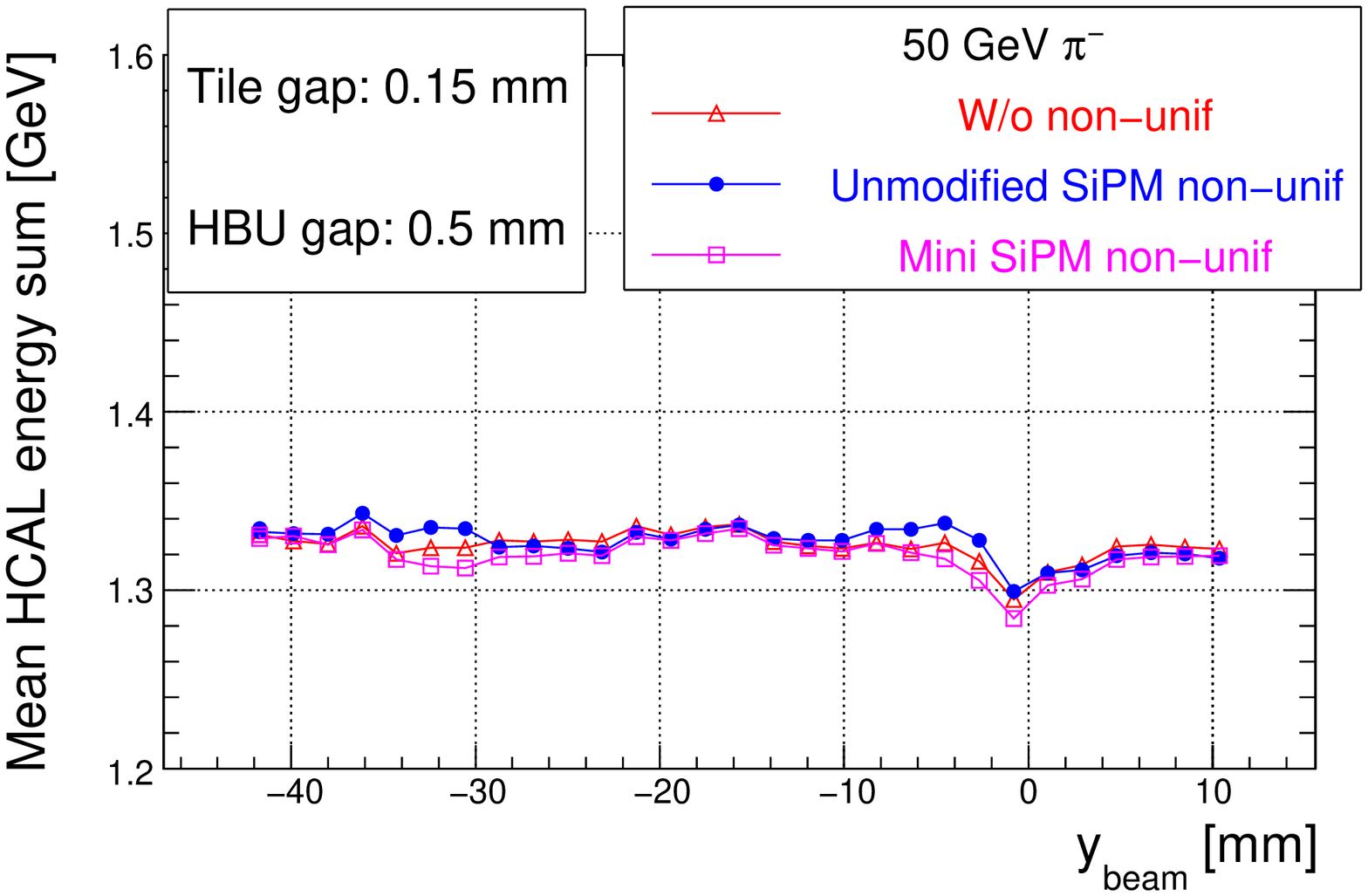}
\caption{Mean HCAL energy sum as a function of the beam position along the $x$ (top) and the $y$-axis (bottom), respectively, in case of \textbf{realistic} gaps between tiles \textbf{without wavelength shifting fiber}, for 50 GeV electrons (left) and pions (right).}
\end{center}
\label{fig:scanXY-27.5}
\end{figure}

\subsection{Staggered layers}

In a possible future linear collider, the HCAL will surround an electromagnetic calorimeter, and in addition, the tiles will be most probably staggered in subsequent layers, and not aligned, as in the test beam case. Since offsets of the order of the tile size ($3\times3\;\textrm{cm}^2$) are expected, the HCAL layers in the test beam simulation were shifted randomly, based on a Gaussian randomisation with mean zero, and a spread of 3 cm. The obtained results are shown in Fig~7. When the layers are staggered, the small remaining effects due to gaps, in the pion  case, cancel away.

\begin{figure}[h!]
\begin{center}
\includegraphics[scale=0.3]{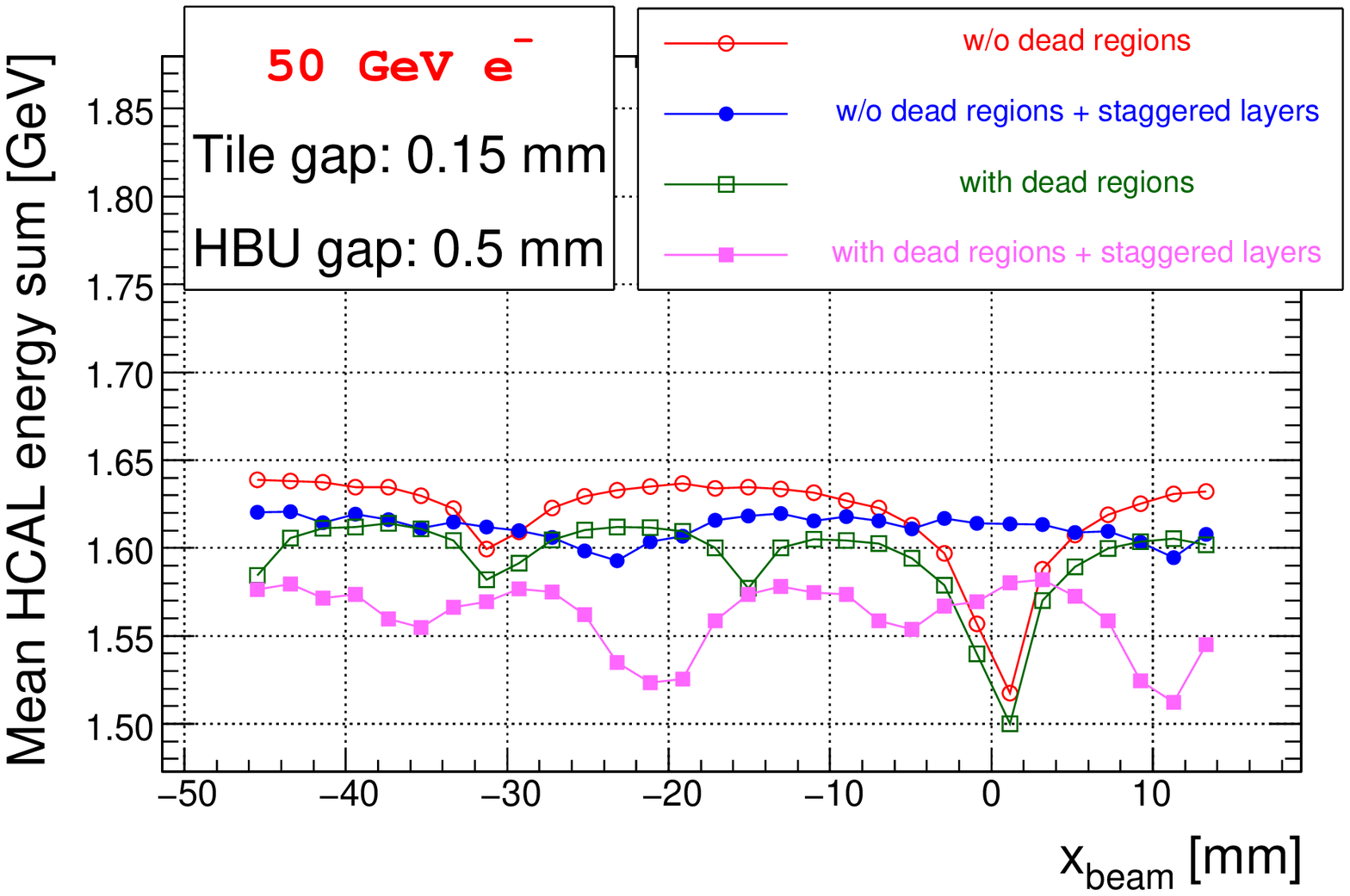}
\includegraphics[scale=0.3]{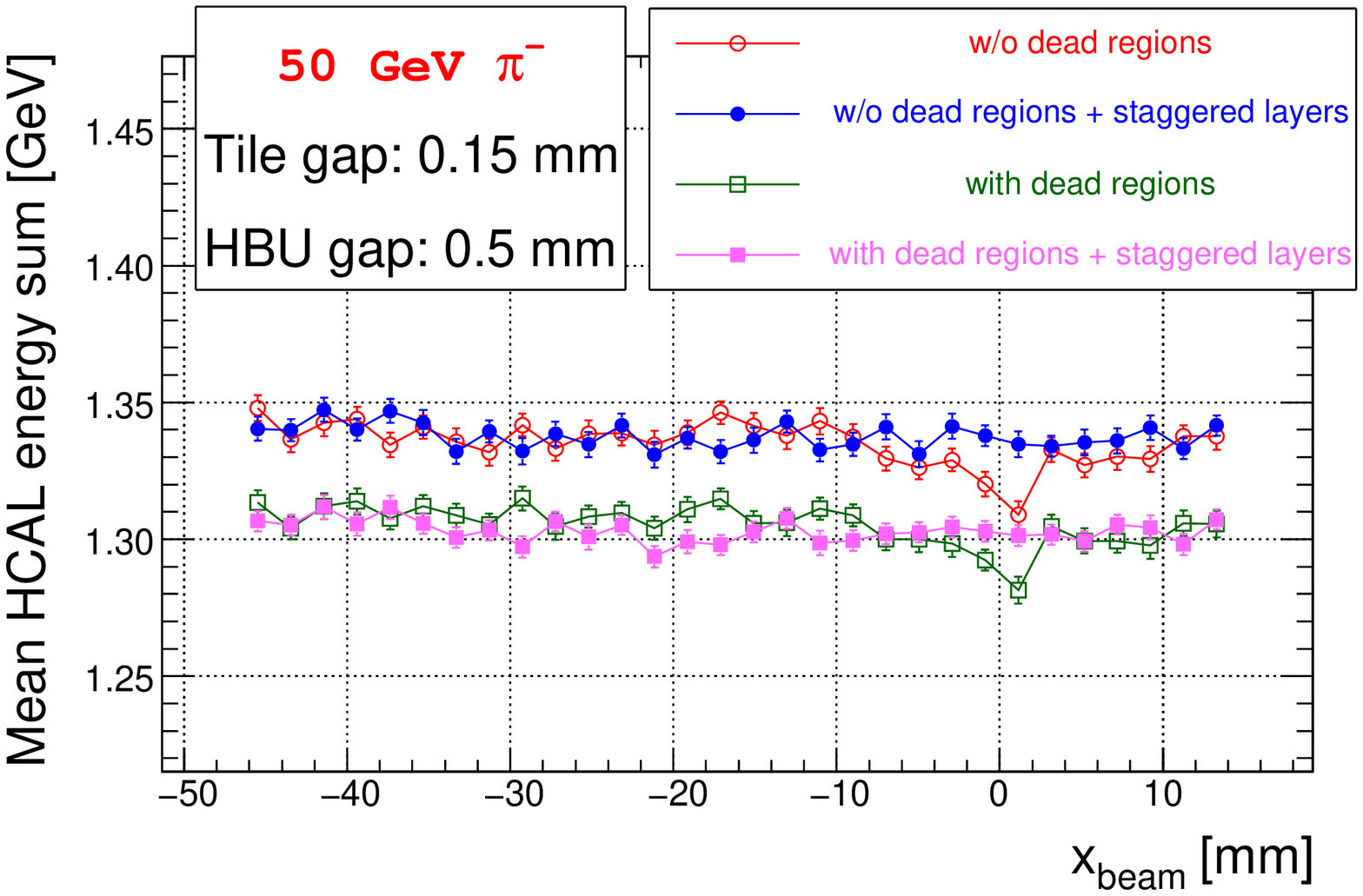}
\caption{Mean HCAL energy sum as a function of the beam position along the $x$ (top) and the $y$-axis (bottom), respectively, in case of \textbf{realistic} gaps between tiles \textbf{without wavelength shifting fiber}, for 50 GeV electrons (left) and pions (right).}
\end{center}
\label{fig:stagger}
\end{figure}

\section{Conclusions}

The effects of gaps between the scintillator tiles of the CALICE hadronic calorimeter, and of the tiles' non-uniformity on the total energy deposited in the calorimeter were studied with simulated events. The gaps and the non-uniformity are not simulated by default, due to technical reasons: the level of complexity would increase very much, since there are several millions of tiles which would need to be simulated separately, and due to GEANT4 limitations the simulation of the physics processes through those tiles would be slown down significantly. But the most important reason for not simulating the gaps and the tiles non-uniformities is that we expected them not to have any significant impact on the physics results. This was proven to be the case in the presented simulation studies.


\begin{footnotesize}


\end{footnotesize}



\begin{thebibliography}{99}
\bibitem{Hcal-in-Mokka-ILD-note} A.~Lucaci-Timoce and R.~Diener, {\it Description of the HCAL Detector in Mokka}, LC-Note  {\bf LC-TOOL-2008-001}, http://www-flc.desy.de/lcnotes/notes/LC-TOOL-2008-001.pdf

\bibitem{EUDET-report} E.~Garutti {\it et~al.}, {\it HCAL Mechanical Design and
Electronics Integration}, {\bf EUDET-Report-2009-07}

\bibitem{ECAL-paper} 
  C.~Adloff {\it et al.}  [CALICE Collaboration],
{\it Response of the CALICE Si-W Electromagnetic Calorimeter Physics Prototype
  to Electrons}, 
  J.\ Phys.\ Conf.\ Ser.\  {\bf 160}, 012065 (2009)
  [arXiv:0811.2354 [physics.ins-det]].

\bibitem{Soldner} C. Soldner, {\it Scintillator Tile Uniformity Studies
for a Highly Granular Hadron Calorimeter}, diploma thesis, Max-Planck-Institut f\"ur Physik, Sept. 2009

\end{thebibliography}
\end{document}